\documentclass[prl,10pt, preprint, nofootinbib, longbibliography, letterpaper]{mynature}

\usepackage{graphicx}
\usepackage{ae,aecompl}
\usepackage{bm}
\usepackage{amsmath}
\usepackage{amssymb}
\usepackage{textcomp}
\usepackage{xcolor}
\usepackage{soul}
\usepackage{bbold}
\usepackage{pdfpages}
\usepackage{lmodern}

\usepackage{anysize}
\marginsize{1.4cm}{1.4cm}{2.8cm}{2.3cm}

\setcitestyle{super,longbibliography}

\usepackage[colorinlistoftodos, shadow,textsize= footnotesize]{todonotes}

\newcommand{\vect}[1]{\bm{#1}}
\newcommand{\be}{\begin{equation}}
\newcommand{\ee}{\end{equation}}

\newcommand{\beq}{\begin{eqnarray}}
\newcommand{\eeq}{\end{eqnarray}}

\newcommand{\LENS}{LENS European Laboratory for Nonlinear Spectroscopy, and Dipartimento di Fisica e Astronomia, Universit\`a di Firenze, 50019 Sesto Fiorentino, Italy}
\newcommand{\INO}{Istituto Nazionale di Ottica-CNR, 50019 Sesto Fiorentino, Italy}
\newcommand{\QSTAR}{Quantum Science and Technology in Arcetri, QSTAR, 50125 Firenze, Italy}

\newcommand{\ICFO}{The Institute of Photonic Science, ICFO, Av. Carl Friedrich Gauss 3, 08860 Castelldefeis, Spain}
\newcommand{\BRA}{Instituto de F\'{i}sica de S\~{a}o Carlos, Universidade de S\~{a}o Paulo, C.P. 369, 13560-970, S\~{a}o Carlos, S\~{a}o Paulo, Brazil}

\begin{document}

\title{\begin{flushleft} 
Observation of Quantum Phase Transitions with Parity-Symmetry Breaking and Hysteresis
\end{flushleft}}

\author{\begin{flushleft} \textsf{\textbf{
A. Trenkwalder$^1$, G. Spagnolli$^2$, G. Semeghini$^2$, S. Coop$^{2,3}$, M. Landini$^1$, P. Castilho$^{1,4}$, L. Pezz\`e$^{1,2,5}$, G. Modugno$^2$, M. Inguscio$^{1, 2}$, A. Smerzi$^{1,2,5}$, M. Fattori$^{1,2}$
}} \end{flushleft}}

\affiliation{\flushleft \textsf{
$^1$ \INO \\
$^2$ \LENS \\
$^3$ \ICFO \\
$^4$ \BRA \\
$^5$ \QSTAR \\ 
}}

\maketitle

{\bf 
Symmetry-breaking quantum phase transitions play a key role in several condensed matter, cosmology and nuclear physics theoretical models \cite{Vojta, review}. 
Its observation in real systems is often hampered by finite temperatures and limited control of the system parameters. 
In this work we report for the first time the experimental observation of the full quantum phase diagram across a transition where the spatial parity symmetry is broken.
 Our system is made of an ultra-cold gas with tunable attractive interactions trapped in a spatially symmetric double-well potential. 
 At a critical value of the interaction strength, we observe a continuous quantum phase transition where the gas spontaneously localizes in one well or the other, thus breaking the underlying symmetry of the system. Furthermore, we show the robustness of the asymmetric state against controlled energy mismatch between the two wells. 
 This is the result of hysteresis associated with an additional discontinuous quantum phase transition that we fully characterize. 
Our results pave the way to the study of quantum critical phenomena at finite temperature\cite{Sachdev}, 
the investigation of macroscopic quantum tunneling of the order parameter in the hysteretic regime and the production of strongly quantum entangled states at critical points\cite{Fazio}.}

\newpage

Parity is a fundamental discrete symmetry of nature \cite{Feynman}, conserved by gravitational, electromagnetic and strong interactions \cite{LeeYang}. 
It states the invariance of a physical phenomenon under mirror reflection. 
Our world is pervaded by robust discrete asymmetries, spanning from the imbalance of matter and antimatter to the homo-chirality of DNA of all living organisms \cite{Pasteur}.
Their origin and stability is a subject of active debate. Quantum mechanics predicts that asymmetric states can be the result of phase transitions occurring at zero temperature, named in the literature as quantum phase transitions (QPTs) \cite{Sachdev, review, Vojta}. The breaking of a discrete symmetry via a QPT provides also asymmetric states that are particularly robust against external perturbations. Indeed, the order parameter of a continuous symmetry breaking QPT can freely (with no energy cost) wander along the valley of a ''mexican hat'' Ginzburg-Landau potential (GLP) by coupling with gapless Goldstone modes \cite{Goldstone}. In contrast, the order parameter of discrete symmetry-breaking QPTs is governed by a one dimensional double-well GLP \cite{Landau}. The reduced dimensionality suppresses Goldstone excitations and the order parameter can remain trapped at the bottom of one of the two wells. This provides a robust hysteresis associated with a first order QPT.

Evidences of parity symmetry breaking has been reported in relativistic heavy-ions collisions \cite{Abelev} and in engineered photonic crystal fibers \cite{Yacomotti}. Observation of parity symmetry breaking in a QPT has been reported for neutral atoms coupled to a high-finesse optical cavity \cite{Esslinger}. However this a strongly dissipative system with no direct access to the symmetry breaking mechanism necessary to study the robustness of asymmetric states. In addition, previous theoretical studies \cite{Claverie, Lasinio} have interpreted the puzzling spectral properties of a gas of pyramidal molecules that date back to the 50s \cite{Loubster}, in terms of the occurence of a QPT with parity symmetry breaking.

In the present work we report the observation of the full phase diagram across a QPT where the spatial parity symmetry is broken. Our system consists of ultra-cold atoms trapped in a double-well potential \cite{Oberthaler, Schmiedmayer} where the tunable strength of the attractive interparticle interaction is the control parameter of the transition. 
Additional control of the energy mismatch between the two wells allows driving of discontinuous 
first order quantum phase transitions in the non-symmetric ordered part of the phase diagram and observation of an associated hysteretic behavior.  

In our system the atomic ground state depends on two competing energy terms in the Hamiltonian $H = H_a + g H_b$, 
where $H_a = \int d\vect{r} \, \Psi^\dag(\vect{r}) [-\frac{\hbar^2}{2m}\nabla^2 + V(\vect{r})] \Psi(\vect{r})$ includes kinetic and potential energy,  
$H_b = \tfrac{2\pi \hbar^2 a_0}{m} \int d\vect{r} \, \Psi^\dag(\vect{r}) \Psi^\dag(\vect{r}) \Psi(\vect{r}) \Psi(\vect{r})$ accounts for contact interaction between the atoms.
Here, $\Psi(\vect{r})$ is the many-body wave function (in the following we consider normalization 
$\langle \Psi^\dag(\vect{r})\Psi(\vect{r}) \rangle =1$),  $m$ the atomic mass, $a_0$ is the Bohr radius
and $V(\vect{r})$ is a double-well trapping potential in the $x$ direction, see Fig.~1a, and a harmonic trap in the orthogonal plane. 
The adimensional control parameter $g = N a_s/a_0 < 0$ is the product of
the total number of atoms $N$ and the scattering length $a_s < 0$ characterizing the interatomic attractive interaction.
The full many-body Hamiltonian is invariant under $x \leftrightarrow -x$  mirror reflection. 
This parity symmetry imposes a spatially symmetric ground state for any value of the control parameter $g$. 
Since $H_a$ and $H_b$ do not commute, the corresponding ground states are quite different.
$H_a$ is minimized by each atom equally spreading on both wells.
A finite energy gap, specified as tunneling energy $J$, separates the ground  
and the first (antisymmetric) excited state of $H_a$. 
$J$ can be tuned by controlling the height of the potential barrier between the two spatial wells.
In contrast, $g H_b$ is minimized by a linear combination of two degenerate states, one having all atoms localized 
in one well, the second with all atoms localized in the other well. 
Because of the competition between the two terms in the Hamiltonian, 
the energy gap between the two low-lying states vanishes (strictly equal to zero in the thermodynamic limit $N \to \infty$, $a_s \to 0$)
at a finite critical value of the control parameter $g_c$.
For $g>g_c$ the tunneling energy dominates and the system is in a symmetric configuration. 
For $g \leq g_c$, when interactions prevail, the system becomes exponentially sensitive to arbitrarily 
small fluctuations of the energy of the two wells. 
This forces the majority of atoms to localize randomly in one well or the other, see Fig~1a.
The broken symmetry is characterized by a non-zero order parameter that, in our case, is
the normalized atomic population imbalance $z = (N_L - N_R)/N$,
where $N_L$ and $N_R$ are the number of atoms occupying the left and the right well, respectively. 

In order to measure $z$ across the phase transition (see Fig.~1b) we adopt the following experimental procedure:
We start by cooling a gas of $N=4500$ atoms well below the Bose-Einstein condensation point until no thermal fraction can be detected (see Methods).
The atoms are initially trapped in a harmonic potential, with a positive scattering length $a_s = 3 a_0$. 
We reach different target values of $g$, above and below the critical point, by continuously transforming
the harmonic trap into a double well, up to a certain barrier height and tunneling $J \approx 40$ Hz, and by 
tuning the interatomic scattering length to negative values (see Methods). 
When $g > g_c$, we find the system in a parity-symmetric disordered phase, with order parameter $z \approx 0$ within error-bars. 
Below the critical value, i.e. for $g < g_c$, an ordered phase emerges with $z$ driven away from zero. 
The phase transition can be theoretically described by an effective GLP 
$W(z) =  \frac{\tilde{g}}{2} z^2 -  \sqrt{1-z^2}$, where 
$\tilde{g} = {g U}/{J}$ is the normalized control parameter and $U$ is the bulk energy (see Methods).
When $\tilde{g}$ crosses the critical value $\tilde{g}_c = -1$, the shape of $W(z)$ continuously changes from a parabola to a double well
(see Fig.~1c) with minima located at $z = \pm \sqrt{1-1/\tilde{g}^2}$~\cite{Raghavan}.
The continuous variation of $z$ indicates the occurrence of a continuous phase transition.
From the experimental measurements of the order parameter 
we obtain a critical value $\tilde{g}_c = -1.3 \pm 0.2$, in fair agreement with the theoretical prediction. 

A one-dimensional GLP, which depends on a single real parameter, describes a
phase transition with the breaking of a discrete symmetry as, for instance, the left-right symmetry in our case,
or the spin-up/spin-down symmetry in the paramagnetic-to-ferromagnetic transition \cite{Sachdev}. 
In these systems, a controlled symmetry-breaking term 
(in our case provided by an energy gap $\delta$ between the two wells, see Fig.~2a) 
drives a first-order QPT in the ordered region of the phase diagram.
This can be understood from the sudden variation of the absolute minimum of the GLP $W(z) + \delta z$
when tuning $\delta$ from positive to negative values, as shown in Fig.~2b.
We characterize this QPT, by repeating the previous experimental procedure, above and below the critical point, but adding to the final 
potential configuration a finite and controlled value of $\delta$. 
As shown in Fig. 2c, for $g > g_c$, the order parameter $z$ changes smoothly as a function of $\delta$, 
with a finite susceptibility $\chi=\vert \frac{d z}{d \delta} \vert$ (measured around $z=0$). 
Increasing the strength of the attractive interaction
causes $z$ to depend more and more critically on $\delta$,  and $\chi$ diverges at the critical point.
Performing the susceptibility measurements for $g < g_c$ we 
observe an abrupt jump of $z$ when crossing the value $\delta=0$, signaling the onset of the first-order phase transition.

The full phase diagram where the interplay of the observed discontinuous (first-order) and continuous (second-order) QPTs is summarized in Fig.~4a. 
We identify the universality class of the parity-symmetry breaking QPT from the susceptibility measurement.
A fit of $\chi$ as a function of the interaction strength according to $\chi = \alpha / (g-g_c)^{\gamma}$ is shown in Fig.~3. 
The two fitting parameters are $\gamma$ and $\alpha$. 
We obtain a critical exponent $\gamma = 1.0 \pm 0.1$, in agreement with the Ginzburg-Landau prediction $\gamma=1$ (see Methods). 
Therefore, our QPT belongs to the universality class of the Lipkin-Meshkov-Glick \cite{LMG} model. 

Discrete-symmetry models are characterized by metastability and hysteresis when driving the system across the first-order transition. 
Both follow directly from the 1D nature of the effective GLP. 
We notice that bifurcation and hysteresis are typical phenomena in the dynamics of 
systems governed by a nonlinear equation of motion \cite{Raghavan, Oberthaler_bif, Phillips}.  
In our system, the origin of hysteresis can be understood from the shape of the GLP $W(z) + \delta z$, which, for 
$\tilde{g}<-1$ and $|\delta| < \delta_c = [(-\tilde{g})^{2/3}-1]^{3/2}$ shows 
an absolute minimum and a local one, see Fig.~4b and Methods.
The latter corresponds to a metastable point with a  
a lifetime depending on the macroscopic quantum tunneling rate of the order parameter through the effective GLP.
This rate is exponentially smaller than the interwell tunneling rate of the single atoms in the double-well trap.
To demonstrate hysteresis in our system, we set $J\approx 30$ Hz, add an energy gap $\delta_0$ 
and prepare a condensate in the well with lower energy 
(for example the right one, with $\delta_0 = -4 J < 0$ and $z \approx 1$). 
We then shift the relative energy of the two wells to a final value $\delta$ in 500 ms, keeping $J$ constant, and 
measure the order parameter after a short waiting time of 10 ms (green circles in Fig.~4c). 
The experiment is performed for different values of the control parameter $g$.
When $g < g_c$, the strong attractive interaction between atoms forces the condensate to 
remain localized in the right well even when its energy minimum is lifted above the left well. 
When $\delta$ overcomes a critical value $\delta_c$, a spinoidal instability drives the gas 
down to the left toward the absolute minimum of the trapping potential
in a timescale $\approx 10$ ms, a fraction of $1/J$.
An analogous behavior is observed with an initial imbalance $z \approx -1$ (orange circles in Fig.~4c), forming an hysteresis loop.
The area of the hysteresis loop decreases with
increasing $g$ and disappears for $g > g_c$, see Fig.~4c.

This work paves the way to the study of macroscopic quantum tunneling in the hysteretic 
regime in the context of the quantum-to-classical transition problem \cite{Zurek}.
Furthermore it will also be possible to explore quantum criticality at finite temperature as consequence of the competition 
between thermal fluctuations and quantum correlations \cite{Copp}.
In addition it will be interesting to explore spontaneous symmetry breaking in gas mixtures as a function of the interspecies interactions \cite{Presilla}.
Finally our system will allow investigation of the creation of entanglement at the critical points \cite{Fazio} 
as a resource for precision measurements \cite{Pezze} and other quantum technologies \cite{Giovannetti}. \\


\begin{figure}[ht]
\begin{center}
\includegraphics[clip,scale=0.44] {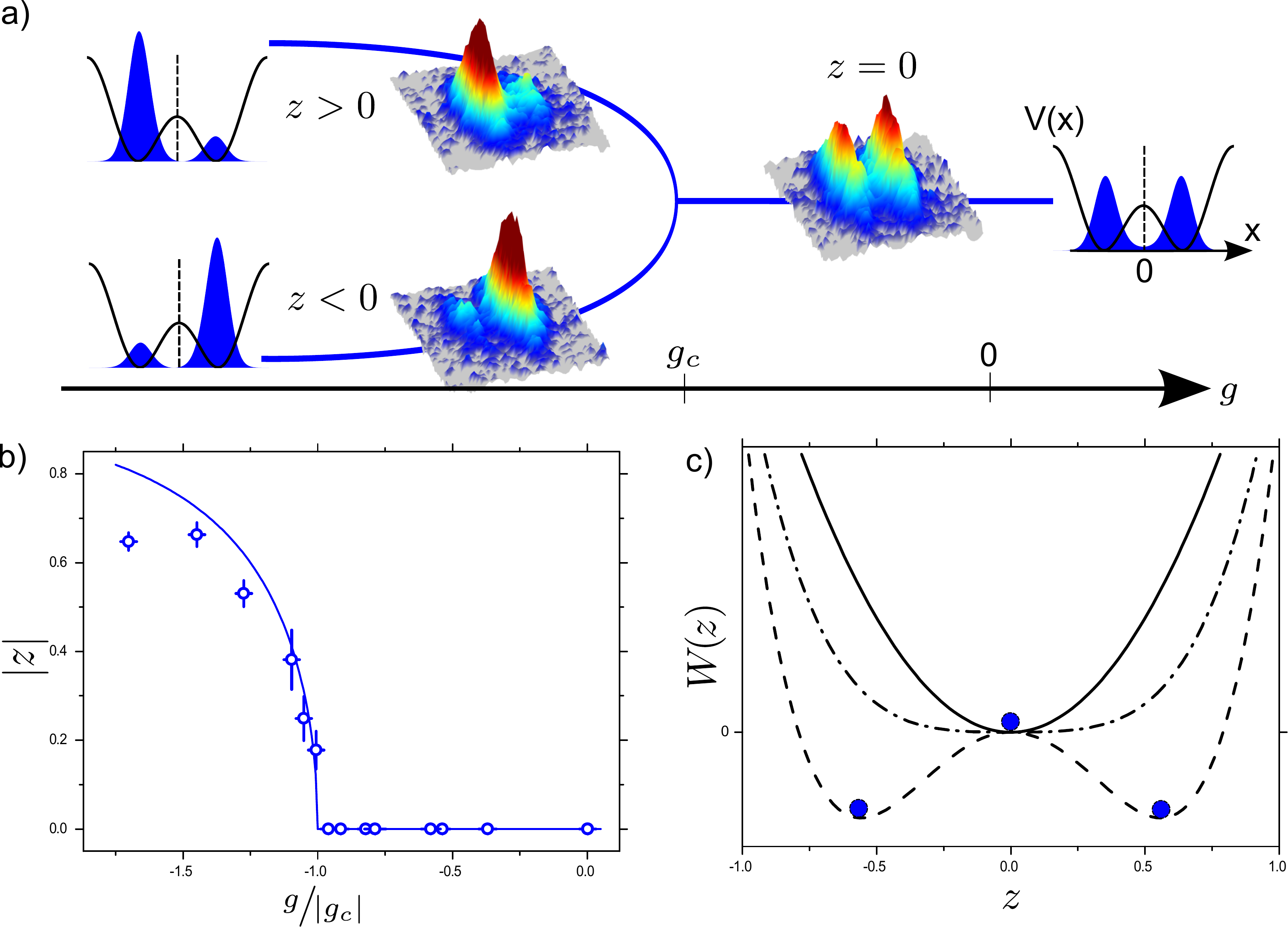}
\end{center}
\caption{\textbf{Schematics of the parity-symmetry-breaking QPT.} 
a) The ultracold atomic gas (blue wavefunction) is trapped in a double-well potential (black line). 
Tuning the interatomic interaction strength $g$ to large negative values, the ground state of the system 
goes from a gapped symmetric state (atomic imbalance $z = \vert (N_L - N_R)/N \vert =0$) to two degenerate asymmetric states ($\vert z \vert >0$). 
The system undergoes a second-order QPT where the spatial parity symmetry, i.e. reflection with respect to the vertical dotted line (symmetry axis), is broken. 
We show experimental absorption images in pseudo 3D and false colors of the atoms obtained at different values of $g$.  
b) Absolute value of the order parameter $z$ as a function of the control parameter $g$ in a balanced double well. 
Error bars are 3 times the standard deviation (see Methods).  
The solid line is the fitting function with $z = 0$ for $g > g_c$ and $z = \sqrt{1 - (\tfrac{g_c}{g})^2}$ for $g \leq g_c$.
From a fit to the data we extract $g_c$ with a relative uncertainty of 0.15. 
It agrees within 20\% with the theoretical prediction.   
c) GLP $W(z)$ for different values of $g/|g_c|$ across the QPT:
$g/|g_c| = - 0.5 $, solid line; $g/|g_c| = -1$, dash-dotted line; $g/|g_c| = - 2$, dashed line. 
}       \label{2DMOT}
\label{efficiency}
\end{figure}

\newpage 


\begin{figure}[ht]
\begin{center}
\includegraphics[clip,scale=0.28] {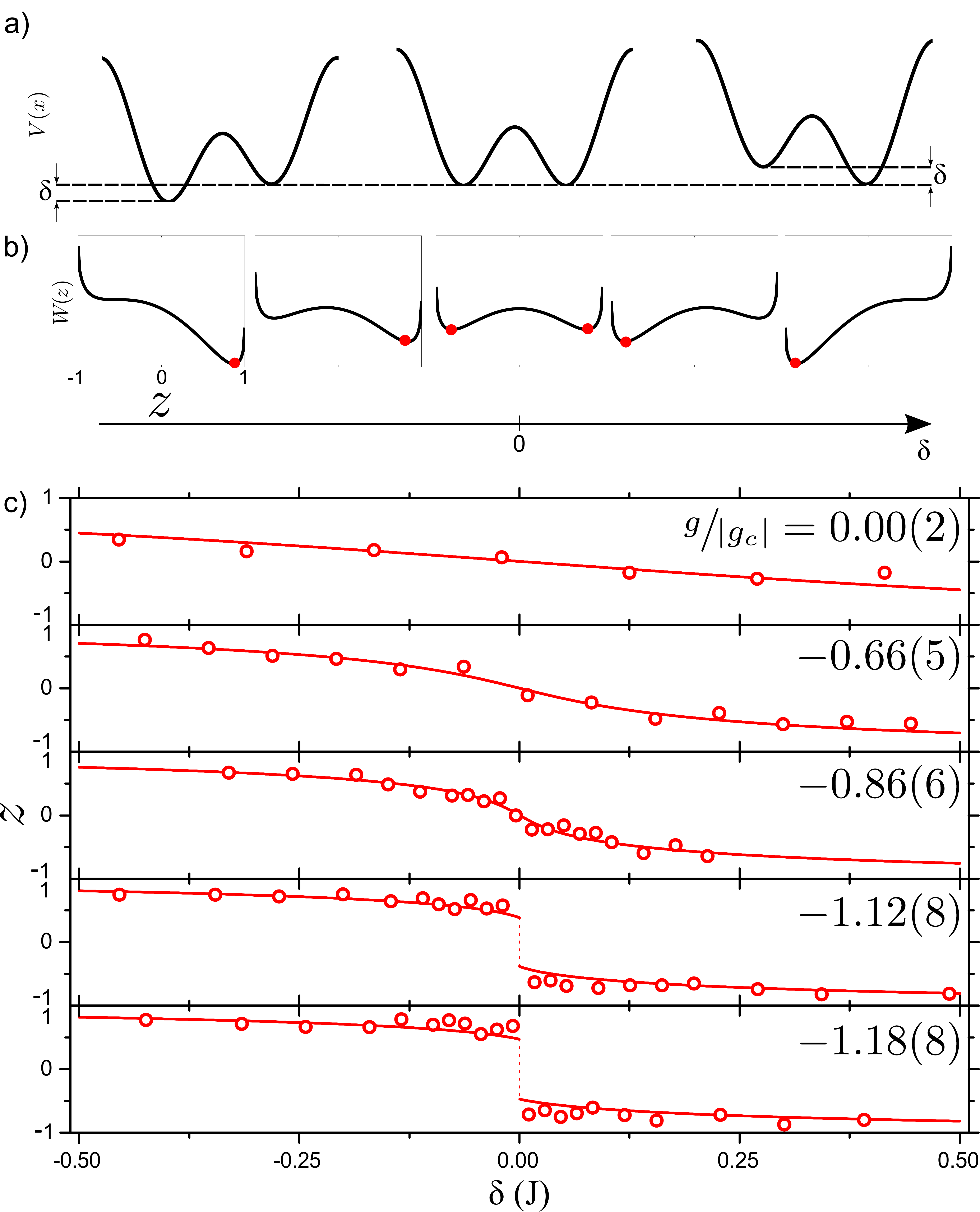}
\end{center}
\caption{\textbf{First-order QPT.} 
a) Our full control of the double-well potential include the tuning of the energy gap $\delta$ between the two minima (see Methods).
b) GLP $W(z)+\delta z$, plotted for $g < g_c$. 
Red dots represents the ground state of the system. 
c) Ground state atomic imbalance $z$ as a function of $\delta$ (circles). 
Different panels correspond to different values of $g / |g_c|$.
For $g \geq g_c$ the atomic imbalance $z$ changes continuously as a function of $\delta$, while for $g < g_c$ the order parameter shows a discontinuity
from positive to negative values signaling the onset of a first-order phase transition. 
The red lines are the result of a fit using the Ginzburg Landau theoretical model (see Methods) with $g_c$ the only fitting parameter. 
 } 
\label{efficiency}
\end{figure}

\newpage


 \begin{figure}[ht]
\begin{center}
\includegraphics[clip,scale=0.6] {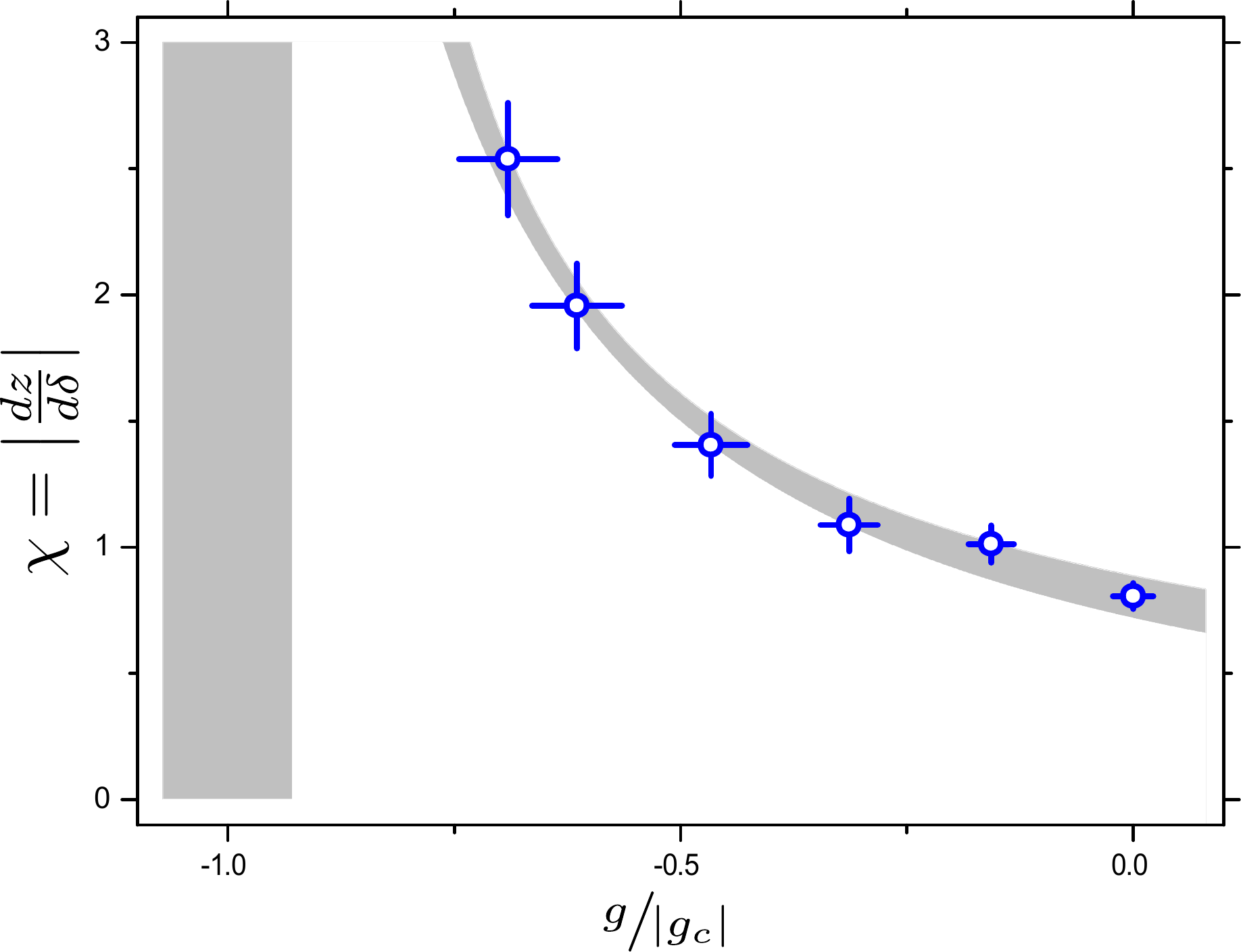}
\end{center}
\caption{\textbf{Susceptibility and measurement of the critical exponent.} 
Susceptibility $\chi= \vert \frac{d z}{d \delta} \vert$ of the system to potential energy gap $\delta$ between the two wells. 
The measurement is performed close to the critical point for values $g > g_c$. 
The curve is a fit to the data (see text)
providing a critical exponent $-1.0 \pm 0.1$ in excellent agreement with the theoretical prediction, equal to -1.
The error bar in the critical exponent comes from the indetermination of $g_c$ (grey region). 
}       \label{2DMOT}
\label{efficiency}
\end{figure}

\newpage


\begin{figure}[ht]
\begin{center}
\includegraphics[clip,scale=0.18] {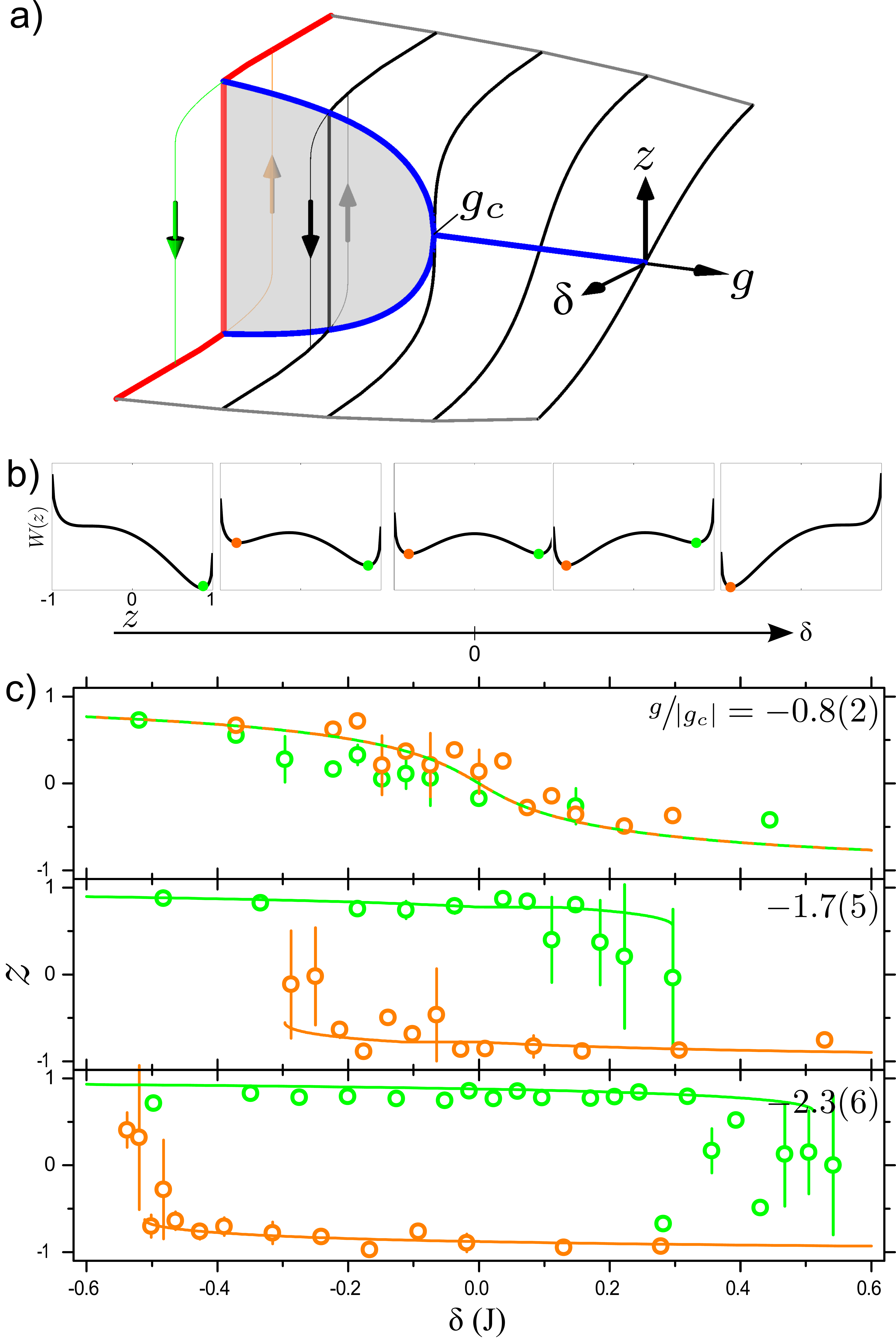}
\end{center}
\caption{\textbf{Full phase diagram and hysteresis.} 
a) Full theoretical phase diagram showing the interplay of the continuos (second-order -- blue line) and discontinuous 
(first-order -- red line) QPTs. 
The green and orange lines shows metastable states giving access to hysteresis.
b) GLP for $g < g_c$ (black line) as a function of $\delta$.
It shows absolute and local minima corresponding to ground and metastable states, respectively.
Orange and green circles represent the states measured in panel (c). 
c) Atomic imbalance as a function of the energy gap $\delta$ between the two wells.
Green (orange) dots are obtained cooling the gas to its ground state at negative (positive) $\delta$ and then 
increasing (decreasing) $\delta$ to the final value in 500 ms and waiting 10 ms before the measurement of the imbalance. 
Lines are theoretical predictions for the imbalance of the ground and the metastable states using the Ginzburg Landau model (see Methods).
Different panels correspond to different values of $g / |g_c|$.
Large error bars signify the dynamical breaking of metastability corresponding to oscillation of the gas between the two wells.
} \label{2DMOT}
\label{efficiency}
\end{figure}


\begin{figure}[ht]
\includegraphics[clip,scale=0.11] {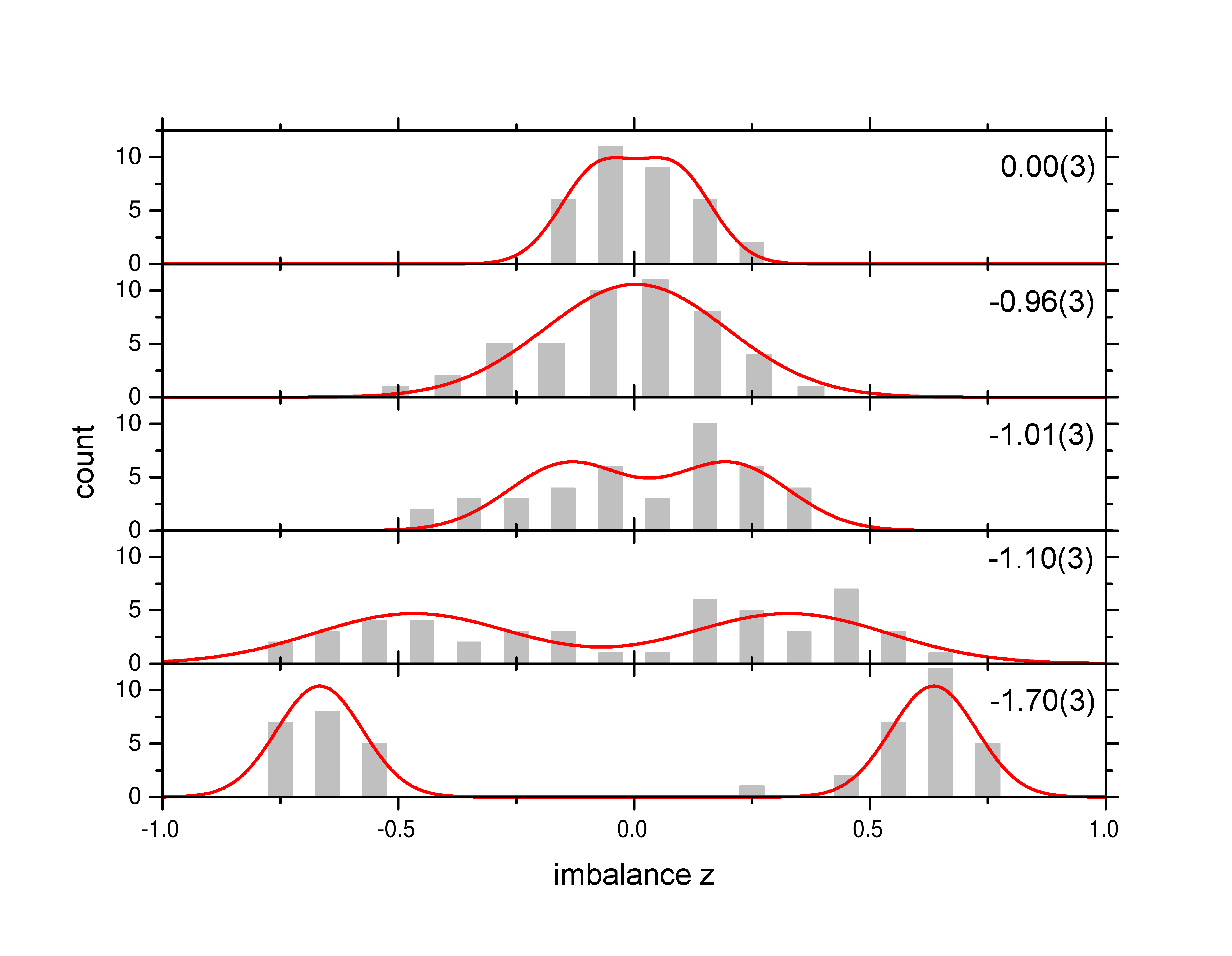}
\caption{\textbf{Histograms of the atomic imbalance } 
The histograms are the result of 40 to 50 measurements of the atomic imbalance z for five different values of the control parameter $g$ with a bin size of 0.1. The numerical values reported on the figure indicate the ratio $g/g_c$. The red curves are the results of a double Gaussian fit described in Methods.     
}\label{2DMOT}
\label{efficiency}
\end{figure}

\section{Methods}

{\bf Double-well potential.}
The double-well potential is created by intersecting at a small angle of 0.1 radiants two pairs of laser beams of 
wavelengths $\lambda$ = 1064\,nm and $\lambda/2$ = 532\,nm which propagate almost perfectly collinearly. 
The interference of the four lasers creates in the horizontal $x$ direction a primary lattice and a secondary lattice with 10 $\mu$m and 5 $\mu$m spacing, respectively. 
Together they form an array of double-well traps \cite{Porto}. Note that the resonance wavelength for $^{39}$K is 767\,nm. 
During the evaporation \cite{Landini} a ``cross'' beam ($\sim 25\,\mu$m waist, 1070\,nm wavelength) allows to load a Bose Einstein condensate (BEC) in a single site of the primary lattice. 
Slowly increasing the intensity of the secondary lattice it is possible to convert the initial harmonic trapping potential into a double-well trap. 
Controlling the relative phase and frequency of the two wavelengths it is possible to control the position of the barrier and eventually introduce an energy gap $\delta$ between the two wells.  
The experiments regarding the second-order phase transition are performed with a primary (secondary) 
lattice depth of 26 (19) nK which provides a tunneling energy of 40 Hz.
The measurements reported in Fig. 2 have been performed starting with a condensate  with slight repulsive interactions initially trapped in a single site of the primary lattice.
In order to set the final target value of $J$ and $g$, we adopt the following experimental procedure. 
In 100 ms we tune the scattering length to the final desired negative value (thus setting $g$). 
We then rise in 500 ms the secondary lattice continuously transforming
the harmonic trap into a double well (thus setting $J$). 
The timescale of the sequence is chosen so to minimize the excitations when the system crosses the transition, 
but short enough to neglect three-body loss processes. 

{\bf BEC with tunable interactions.}
The experiments are performed with atoms in the $F=1, M_F=1$ internal state across a magnetic Feshbach resonance at 402 Gauss~\cite{Derrico}. 
The magnetic-field stability allows a tuning of the scattering length better than 0.1 $a_0$. 
Condensates of 4600$\pm$400 atoms with a scattering length of 3 $a_0$ form in a single site of the primary 
lattice with a longitudinal trapping frequency of 160 Hz. 
An additional radial beam at 1064\,nm, collinear with the lattice, provides a radial confinement of $\omega = 180$ Hz. 
The temperature of the gas is a fraction of the lattice depth and is then well below the condensation temperature (T$_C \approx$ 130 nK).   

{\bf Parity symmetry-breaking QPT measurement.}
For the measurement reported in Fig. 1b we take 40 to 50 images for each datapoint and we measure the imbalance $z$.
For each datapoint we collect a histogram 
with bins of varying bin sizes $\delta z = 0.01 \ldots 0.20$ and $z$ in the range from $-1$ to $+1$ (see Fig. 5). 
We fit each histogram with the sum of two Gaussian distributions:
\begin{equation} 
n(z) = n_{peak} \left(\exp\left[-\frac{(z-\Delta z)^2}{2\,\sigma^2}\right]+\exp\left[-\frac{(z+\Delta z)^2}{2\,\sigma^2}\right]\right)\ , \nonumber
\end{equation}
with $\sigma$ and $n_{peak}$ the width and height of both Gaussians and 2$\Delta z$ the separations between the two Gaussians. 
In Fig.~2a, we plot for each interaction strength, the average of $\Delta z$ over all bin sizes. Errorbars is the standard deviation over all bin sizes. 

{\bf Theoretical Model.}
To theoretically model our system we consider a two-mode approach.  
We approximate the many-body wave function as 
$\Psi(\vect{r}) = \sqrt{\tfrac{N_L}{N}} \psi_L(\vect{r}) e^{i \phi_L} + \sqrt{\tfrac{N_R}{N}} \psi_R(\vect{r}) e^{i \phi_R} $.
Here, the wave functions for the left and right well are identified as
$\psi_L(\vect{r}) = \tfrac{\psi_{\rm gs}(\vect{r}) + \psi_{\rm ex}(\vect{r}) }{\sqrt{2}}$ and 
$\psi_R(\vect{r}) = \tfrac{\psi_{\rm gs}(\vect{r}) - \psi_{\rm ex}(\vect{r}) }{\sqrt{2}}$, respectively,
where $\psi_{\rm gs}(\vect{r})$ and $\psi_{\rm ex}(\vect{r})$ are the ground state, of energy $E_{\rm gs}$, 
and first excited state, of energy $E_{ex}$, of the (single-particle)
Schr\"odinger equation in the symmetric double-well potential.
We indicate as $J = E_{\rm ex} - E_{\rm gs}$ the tunneling energy and as
$U = \tfrac{4 \pi \hbar^2 a_0}{m} \int d \vect{r} \vert \psi_{L,R}(\vect{r}) \vert^4$ the bulk energy. 
After some algebra we obtain \cite{Raghavan}
$H(z, \phi) = \tfrac{\tilde{g} z^2}{2} - \sqrt{1-z^2} \cos \phi$, where 
$z = (N_L - N_R)/N$, 
$\phi = \phi_L - \phi_R$ is the phase difference, and
$\tilde{g}$ is the control parameter normalized to $J/U$.
For attractive interactions, phase fluctuations are small and we thus write
$H(z, \phi) = \tfrac{1}{2} \sqrt{1-z^2} \phi^2 + \tfrac{\tilde{g} z^2}{2} - \sqrt{1-z^2}$.
We can perform a quantization of this Hamiltonian by replacing the conjugate variables $z$ and $\phi$
with operators obeying the commutation relation $[z,\phi] = \tfrac{2i}{N}$.
Rewriting $\sqrt{1-z^2} \phi^2 = \phi \sqrt{1-z^2} \phi$ and choosing the $z$ representation $\phi = - \tfrac{2i}{N} \tfrac{d}{d z}$, 
we obtain the one-dimensional Hermitian Hamiltonian $H_{\rm eff} = - \tfrac{2}{N^2} \tfrac{d}{dz} \sqrt{1 - z^2} \tfrac{d }{d z} + W(z)$ for a particle
with $z$-dependent effective mass moving in a GL potential $W(z) = \tfrac{\tilde{g} z^2}{2} - \sqrt{1-z^2}$.
Note that there are alternative approaches \cite{Trippenbach} leading to the same effective Hamiltonian $H_{\rm eff}$. 
In the thermodynamic limit ($N \to \infty$, $a_s \to 0$, with $g = N a_s/a_0$ finite), the effective mass diverges and 
the low lying eigenstates of $H_{\rm eff}$ are strongly localized in the vicinity of the minima of the GL potential. 
A Taylor expansion around $z=0$ gives
$W(z) = \tfrac{(\tilde{g}+1) z^2}{2} + \tfrac{z^4}{8}$.
The quadratic potential for $\tilde{g}> -1$ becomes quartic when $\tilde{g}= -1$, $z_0=0$ being its absolute minimum.
For $\tilde{g}< -1$ we have a bifurcation and two symmetric minima appear at $z_0 = \pm \sqrt{1 - 1/\tilde{g}^2}$.
The  degeneracy between stable points for positive and negative values of $z$ is lifted by an 
energy gap between the two wells. This is introduced by adding the term $\delta z$ to the GL potential.
The stable and metastable points are searched as the minima of $W(z) + \delta z$, providing the equation 
\be \label{stability} \tag{S.1}
\tilde{g} z_0 + \delta +\frac{z_0}{\sqrt{1-z_0^2}} = 0.
\ee
For $\tilde{g} > -1$, the stable points at small $\delta$ can be found by linearizing Eq.~(\ref{stability}). 
We obtain $z_0 = - \delta/(\tilde{g}+1)$ and thus a susceptibility $\chi = \vert \tfrac{d z_0}{d \delta} \vert = 1/\vert \tilde{g}+1\vert $
showing a divergence at $\tilde{g} = -1$ with critical exponent equal to one.
The metastable states, defining the hysteretic mechanism, can be found by linearizing $W(z)$ around the values of 
$z_0$ defined by Eq.~(\ref{stability}), giving $W(z) \approx  \tfrac{(z-z_0)^2}{2} \big[\tilde{g} + \tfrac{1}{(1-z_0^2)^{3/2}}\big]$.
We can thus find a metastable $z_0$ if and only if the harmonic oscillator frequency $\tilde{g} + \tfrac{1}{(1-z_0^2)^{3/2}}$ is positive, 
i.e. if and only if $\tilde{g} \geq (\tilde{g} + \delta / z_0)^3$.
This conditions, together with Eq.~(\ref{stability}), defines the solid lines in Fig.~4.
These lines extending up to critical values $z_c^{\pm} = \pm \sqrt{1 - \tfrac{1}{(-\tilde{g})^{2/3}}}$ for $\delta_c^{\pm} = \pm [(-\tilde{g})^{2/3}-1]^{3/2}$. 
The extremal points $(z_c^+, \delta_c^+ )$ and $(z_c^-, \delta_c^- )$ are connected 
by unstable fixed points 
corresponding to saddle points of $H(z,\phi)$.

\section{Acknowledgements}

We thank our colleagues of the Ultracold Quantum Gases group in Florence for constant support. We acknowledge discussions with G. Jona-Lasinio and C. Presilla.
This work has been supported by ERC Starting grant AISENS, INFN (Firb RBFR08H058\_001, Micra) and by EU-FP7 (QIBEC). S.C. acknowledges support from the Erasmus Mundus Joint Doctorate programme.  

\end{document}